Article

# Preliminary prediction of the basic reproduction number of the Wuhan novel coronavirus 2019-nCoV

Tao Zhou[1], Quanhui Liu[2,#], Zimo Yang[3], Jingyi Liao[4], Kexin Yang[2], Wei Bai[5], Xin Lu[6,#], Wei Zhang[7,#]

1 Big Data Research Center, University of Electronic Science and Technology of China, Chengdu 611731, China
2 College of Computer Science, Sichuan University, Chengdu 610065, China
3 Beijing AiQiYi Science & Technology Co., Ltd., Beijing 100080, China
4 Shenzhen International Graduate School, Tsinghua University, Shenzhen 518055, China
5 CompleX Lab, University of Electronic Science and Technology of China, Chengdu 611731, China
6 College of Systems Engineering, National University of Defense Technology, Changsha 410073, China
7 West China Biomedical Big Data Center, West China Hospital, Sichuan University, Chengdu 610047, China

**Correspondence**
Quanhui Liu, College of Computer Science, Sichuan University, Chengdu 610065, China. Email: quanhuiliu@scu.edu.cn;
Xin Lu, College of Systems Engineering, National University of Defense Technology, Changsha 410073, China. Email: xin.lu@flowminder.org;
Wei Zhang, West China Biomedical Big Data Center, West China Hospital, Sichuan University, Chengdu 610047, China. Email: zhangwei@wchscu.cn

**Funding**: This work was partially supported by the National Natural Science Foundation of China (Grant Nos. 61433014, 11975071, 71771213, 91846301, 71790615, 71774168), the Fundamental Research Funds for the Central Universities and the Hunan Science and Technology Plan under Grant Nos. 2017RS3040 and 2018JJ1034.

**Abstract**

**Objectives:** To estimate the basic reproduction number of the Wuhan novel coronavirus (2019-nCoV).

**Methods:** Based on the susceptible–exposed–infected–removed (SEIR) compartment model and the assumption that the infectious cases with symptoms occurred before 26 January, 2020 are resulted from free propagation without intervention, we estimate the basic reproduction number of 2019-nCoV according to the reported confirmed cases and suspected cases, as well as the theoretical estimated number of infected cases by other research teams, together with some epidemiological determinants learned from the severe acute respiratory syndrome.

**Results:** The basic reproduction number fall between 2.8 to 3.3 by using the real-time reports on the number of 2019-nCoV infected cases from People's Daily in China and fall between 3.2 and 3.9 on the basis of the predicted number of infected cases from international colleagues. Using a newly reported epidemiological determinants for early 2019-nCoV, the estimated basic reproduction number is in the range [2.2,3.0].

**Conclusions:** The early transmission ability of 2019-nCoV is close to or slightly higher than SARS. It is a controllable disease with moderate-high transmissibility. Timely and effective control measures are needed to suppress the further transmissions.

## Introduction

The transmission of pneumonia associated with the novel coronavirus (2019-nCoV) originated in Wuhan city has not yet been effectively blocked. In the meanwhile, the number of confirmed and suspected cases is increasing rapidly. Estimating the epidemiological determinants of 2019-nCoV is significant and urgent regarding the assessment of epidemic transmissibility, the prediction of future trend of epidemic spreading, as well as the design of control measures. The basic reproduction number is the most important parameter to determine the intrinsic transmissibility, defined as the average number of secondary infectious cases generated by an index case in a completely susceptible population without any interventions [1]. During the outbreak of an epidemic, due to interventions and control measures from government, reaction of personal behaviors (sterilizing, wearing masks, washing hands, reducing contacts, etc.), the depletion of susceptible populations, and the seasonality of transmissibility, the basic reproduction number is generalized to the effective reproduction number, which is defined as the average number of secondary cases generated by an infectious case at time $t$, and is denoted by $R_t$. The epidemic is considered to be under control when $R_t<1$.

We assumed that the infected individuals whose onset time of symptoms no later than January 25, 2020 were resulted from the free propagation, i.e., the transmission was without interventions. Regarding the reports of real-time data of 2019-nCoV situation jointed by the People's Daily in China [2] and DXY.cn (an online community for health care professionals) [3], as well as the estimated number of 2019-nCoV infections from the research group led by the Northeastern University [4] (these two data sources are later abbreviated as People's Daily Reports and Northeastern University Reports), we estimated the basic reproduction number of 2019-nCoV based on the susceptible–exposed–infected–removed (SEIR) compartment model.

## Methods

This article was intended to estimate the basic reproduction number under the situation of free propagation, which was the initial stage of the spread of 2019-nCoV without the interventions. Most Chinese people were aware of the outbreak of 2019-nCoV by the mainstream media after 20 January 2020. The Hubei government released the announcement about strengthening the prevention and control measures against 2019-nCoV, and launched the second-level public health emergency response at 2:40am on 22 January, 2020. Thus, the public awareness and effective interventions were absent when the time was prior to this point. As the median value of the incubation period of SARS was 6.4 days (95% CI 5.2 to 7.7 days) [5], and the 2019-nCoV incubation period was 5.1 days according to a recent report on a few confirmed cases [6], we inferred that the confirmed cases before 26 January, 2020 were infected during the free propagation of 2019-nCoV. Meanwhile, the confirmed cases after 25 January were not suitable for the analyses since the cases were generated in the following days were not during the free propagation.

We used the susceptible–exposed–infected–removed (SEIR) compartment model [7] to characterize the early spreading of 2019-nCoV, where each individual could be in one of the following four states: susceptible (S), exposed (E, being infected but without infectiousness), infected (I, with infectiousness) and removed (R). At each time step (in days in later analyses), a susceptible individual would turn to be an exposed individual with probability $\beta$ if she/he contacts with an infected individual, an exposed individual had a probability $\gamma_1$ to become infected, and an infected would be removed with probability $\gamma_2$. The dynamical process of SEIR thus could be described as

$$\frac{dS(t)}{dt} = -\frac{\beta S(t)I(t)}{N}$$

$$\frac{dE(t)}{dt} = \frac{\beta S(t)I(t)}{N} - \gamma_1 E(t)$$

$$\frac{dI(t)}{dt} = \gamma_1 E(t) - \gamma_2 I(t)$$

$$\frac{dR(t)}{dt} = \gamma_2 I(t)$$

where $S(t)$、$E(t)$、$I(t)$ and $R(t)$ respectively represent the number of individuals in the susceptible, exposed, infectious and recovered states at time $t$, $N$ was the total number of individuals in the system such that $N=S(t)+E(t)+I(t)+R(t)$. The infected population during early transmission was negligible compared with the total population, that was, when $t$ approaches 0, $S(t)$ approaches to $N$. The basic reproduction number could then be approximated as [8]:

$$R_0 = (1+\frac{\lambda}{\gamma_1})(1+\frac{\lambda}{\gamma_2})$$

where $\lambda = \ln Y(t)/t$ was the growth rate of the early exponential growth and $Y(t)$ was the number of infected people with symptom by time $t$. The exposed period and the infection period could be expressed as $T_E = 1/\gamma_1$ and $T_I = 1/\gamma_2$, Generation time could then be approximated as $T_g = T_E + T_I$. Denote by $\rho = T_E/T_g$ the ratio of exposed period to generation time, the basic reproduction number could be rewritten as

$$R_0 = 1 + \lambda T_g + \rho(1-\rho)(\lambda T_g)^2.$$

**Results**

To estimate $R_0$, the parameters $\lambda$, $\rho$ and $T_g$ were needed, where $\lambda$ was determined by $Y(t)$. Below was the detailed discussion about these three parameters.

According to the real-time dynamic data of 2019-nCoV by the People's Daily Reports, there were 1408 confirmed cases and 2032 suspected cases by the time of 23:59 on 25 January 2020. In reality, there should be a certain fraction of infected people with symptoms having not been found [4,9,10]. If we ignored this situation and suppose that there a fraction $q$ of the suspected cases would be confirmed further (on the basis of an early report that 41 of the 59 suspected cases were eventually confirmed, the reference value of $q$ was 41/59=0.695), then the number of cases with symptoms on 25 January was $Y(t) = 2032 \times 0.695 + 1408 = 2820$. Note that, the real number of cases was probably much larger than 2820. We considered this number as the optimistic situation (the lower bound of the number of cases having onset of symptoms). If we took the Northeastern University Reports [4] as the reference, there would be about 4 050 cases on 20 January 2020, and about 12

700 cases on 24 January, 2020. This result was mainly based on the number of confirmed cases in overseas which were exported from Wuhan. Even though there might be a large bias due to the highly limited samples, it should be seriously considered as all other methods were also very preliminary and some reports [9,10] showed similar results to the Northeastern University Reports [4]. Read *et al.* [10] inferred that the actual number of confirmed cases was only 5.1%, including the cases without symptoms. If only 5% of symptomatic infections were detected (more pessimistic than the results of Read's study [10]), then according to the confirmed cases (1 408) by 25 January, the number of cases with symptoms was 28 160, ten times larger than the optimistic one. We used it as the upper bound of $Y(t)$ in sensitivity analysis. We set the date 8 December 2019, the presence time of the first pneumonia of unknown aetiology, as $t=0$. In fact, $t$ might be slightly earlier than 8 December, since the symptoms might have appeared for some time before to the hospital. Hence, the estimated basic reproduction number might be a bit higher by using 8 December 2019 as $t=0$. Based on the above analysis, we mainly used $Y(48)=2\ 820$ from the People's Daily Reports, and $Y(43)=4\ 050$, $Y(47)=12\ 700$ form the Northeastern University Reports [4].

The value of $\rho$ for SARS was in the range of [0.5, 0.8], we took $\rho=0.65$ in the absence of more references [5][11]. In the sensitivity analysis, we would consider $\rho$ from 0.5 to 1. As the formula $\rho(1-\rho)$ was symmetry, the range, i.e., [0.5, 1], covers all possible values of $\rho$. Lipsitch *et al.* [12] showed that the average of $T_g$ is 8.4 days for SARS, while in the early outbreaks $T_g$ was higher (the average value was 10.0 days), and they suggested the sensitivity analysis interval as $T_g \in [8,12]$. Imai *et al.* [13] claimed that the case study reported in Chan's research [6] indicated that 2019-nCoV has a much shorter $T_g$. However, we could not yet obtain any solid estimation about the value of $T_g$ as the number of samples was too small the questionnary survey did not sufficiently cover the interests about genration time. Therefore, we mainly concentrated on $T_g=8.4$ (days) and $T_g=10.0$ (days), and took sensitivity analysis with the same interval [10, 14] as suggested by Lipsitch *et al* [12].

Table 1. The basic reproduction numbers and the corresponding key parameters.

| Data Sources | $T_g$ | $Y(t)$ | $t$ | $R_0$ |
|---|---|---|---|---|
| People's Daily Reports | 8.4 | 2820 | 48 | 2.83 |
| People's Daily Reports | 10.0 | 2820 | 48 | 3.28 |
| Northeastern University Reports | 8.4 | 4050 | 43 | 3.22 |
| Northeastern University Reports | 10.0 | 4050 | 43 | 3.78 |
| Northeastern University Reports | 8.4 | 12700 | 47 | 3.34 |
| Northeastern University Reports | 10.0 | 12700 | 47 | 3.93 |

As shown in Table 1, the basic reproduction number fall between 2.8 to 3.3 based on the People's Daily Reports and fall between 3.2 and 3.9 on the basis of the Northeastern University Reports [4]. The estimated value of $R_0$ by the Reports of the Northeastern University was similar to the value estimated by Read *et al.* [10], which was in the range of [3.6, 4.0]. Our estimated values were higher than the ones ([2.1, 3.5], with a median value 2.6) by Imai *et al.* [13]. In accordance with the currently preliminary estimations, the transmissibility of 2019-nCoV was close to SARS. For example, the basic reproduction number of SARS by Lipsitch *et al.* [12] was [2.2, 3.6], and the average basic reproduction number of SARS by Riely *et al.* [14] was 2.7 (95%CI 2.2 to 3.7), but if considering the superspreading events, this average value rised up to 3.4. The basic reproduction number of SARS by Wallinga and Teunis [15] was from 3.1 to 4.2, which was not lower than our pessimistic result. Therefore, if we rely on the People's Daily Reports, the basic reproduction number of 2019-nCoV was not higher than SARS, or even slightly lower than the basic reproduction number of SARS predicted by some other research teams. Even with the more predictions by the Northeastern University Reports, the basic reproduction number of 2019-nCoV was only slightly higher than SARS. It was a controllable disease with moderate-high transmissibility. Concerning the previous experiences in fighting with SARS, the spreading of 2019-nCoV could be quickly reduced (probably in two or three weeks) through timely and effective control measures by government

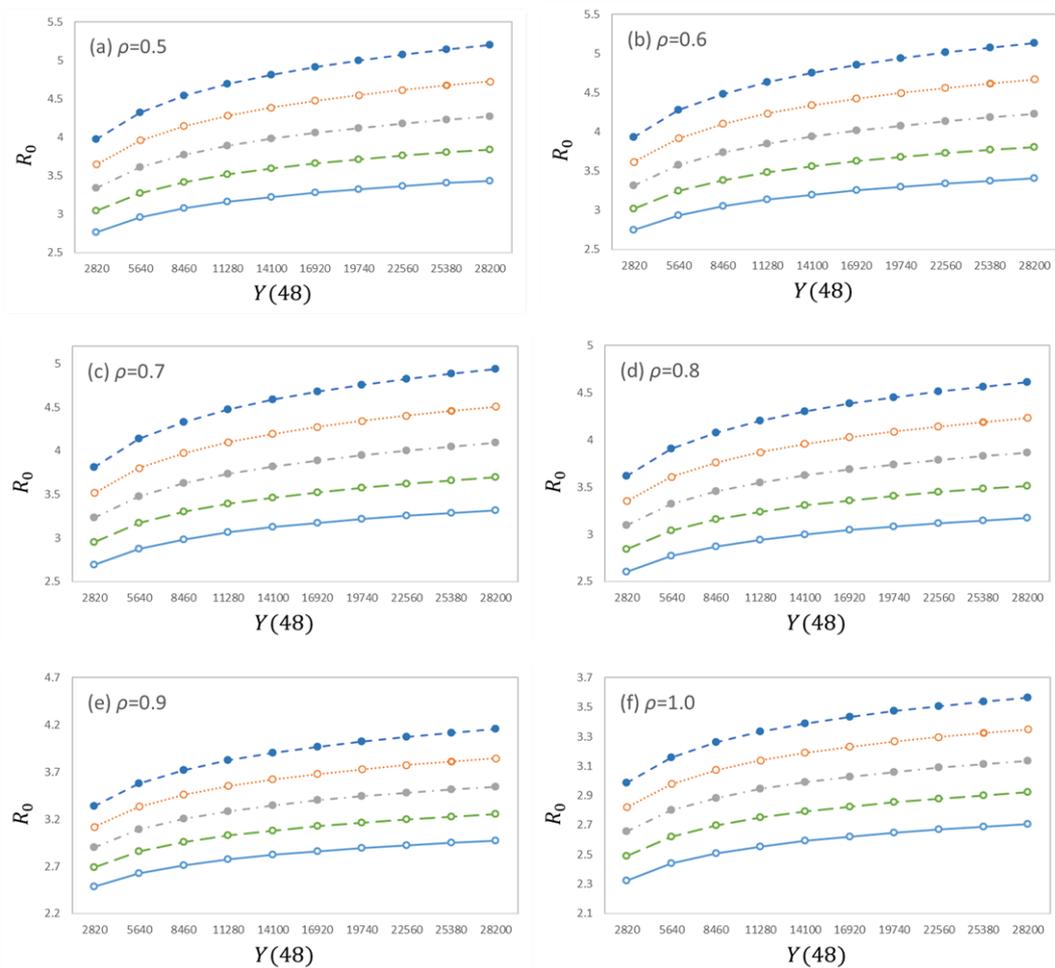

**Figure 1. Impacts of different key parameters on the estimated basic reproduction numbers. The x-axis is the number of cases with symptoms at t=48, figures (a)-(f) correspond to the six different scenarios for $\rho = 0.5, 0.6, 0.7, 0.8, 0.9, 1$, the five curves in each figure from top to bottom correspond to the five different cases for $T_g = 12, 11, 10, 9, 8$.**

Figure 1 showed the sensitivity analysis of the three key parameters. Under the worst case ($Y(48) = 28016, \rho = 0.5, T_g = 12$), the estimated $R_0$ was 5.2; under the best case ($Y(48) = 2820, \rho = 1.0, T_g = 8$), the estimated $R_0$ was 2.3. Because $R_0$ was sensitive to the generation time, with the accumulation of epidemiological survey data, the estimation of generation time would be more accurate, thus providing a more precise estimation of the basic reproduction number.

**Discussion**

Considering the extreme cases, the basic reproduction number was in the range [2.3, 5.2], but we thought it should be in the range [2.8, 3.9]. Based on the data of the People's Daily Reports, the prediction of the basic reproduction number was in the range of 2.8 and 3.3, and this range became [3.2, 3.9] when the data was based on the Reports the Northeastern University [4]. Even with the pessimistic estimation, the basic reproduction number of 2019-nCoV was only slightly higher than SARS, fully equipped with controllable condition. Many known basic reproduction numbers were higher than 2019-nCoV. For example, Zika virus was 1.4-6.6 [16], Middle East respiratory syndrome was 2.0-6.7 [17], and smallpox was 3.5-6.0 [18]. In a word, 2019-nCoV does not have particularly outstanding transmissibility.

The value of basic reproduction number was most sensitive to the generation time $T_g$, and thus we hope the more accurate estimation of the generation time based on the accumulation of epidemiological survey data would further improve the quality of the estimation of $R_0$. The number of confirmed cases obtained from different sources varied widely. Since a significant proportion of patients with 2019-nCoV had mild symptoms, which could be healed without entering the hospital, there might be a large number of patients not in the official confirmed list. Consequently, we suggested to be more prudent. For example, the design of control measure should refer to the more pessimistic estimation (based on the Northeastern University Reports) rather than the more optimistic one (based on the People's Daily Reports). Our model assumes that the individuals in exposed state did not have or have very lower-level infectiousness (according to SARS), however, it was possible that for 2019-nCoV, individuals in the exposed state still had considerable infectiousness. Such possible difference was already taken into account by varying the value of $\rho$.

It was needed to be emphasized that the method used in this article was a preliminary estimation under the premise of largely insufficient data. In order to have a better estimation of the basic reproduction number and effective reproduction number, as well as predicting the trend of epidemic transmission, we not only need to know precise epidemiological determinants, but also need to improve the model itself by further considering the diversity in susceptibilities and contact probabilities of people in different ages and genders, the different spreading mechanisms in hospitals and communities, the effects of regional population density and human mobility, and so on [19-22].

From the dynamic perspective, $R_0 = k\beta D$, where $k$ was the average number of contacts to susceptible individuals of an infected individual per day, $\beta$ was the transmission probability through a contact between an infected individual and a susceptible individual, and $D$ was the effective time period allowing an infected individual to infect susceptible individuals. The government's control policy and individuals' prevention behaviors were to reduce these parameters, and to eventually make the effective reproduction number $R_t$ below 1. According to our results, if $k\beta D$ could be reduced by 3/4 (to its 1/4), 2019-nCoV could be effectively controlled. Staying at home and cancelled meetings could decrease the frequency of contacts between infected and susceptible individuals. Wearing masks and washing hands could reduce the transmission probability $\beta$. If the individuals exhibit the suspected symptoms or have contacts with the high-risk groups, the corresponding medical observation and the isolation with other individuals were needed, which would shorten the effective infectious time period $D$. Restrictions on transportation, extension of winter holiday, and the cancellation of various conferences also aimed at reducing the effective reproductive number. Based on the preliminary information, a considerable amount of people infected with 2019-nCoV only show mild symptoms, and thus they had the same mobility as healthy people in principle. There were also some preliminary epidemiological findings suggesting that infected cases had infectiousness during the exposed period. These reasons, together with the delayed responses by Wuhan government, lead to a fiercer outbreak of 2019-nCoV than SARS. On the contrary, reducing the traveling and avoiding the meetings, in the meanwhile, wearing masks and washing hands frequently would suppress the effects caused by mild symptoms and exposed state with infectiousness. Relying on the Chinese experiences in fighting with SARS, we believed that 2019-nCoV would be effectively controlled soon.

In conclusion, the SEIR model is employed to describe the dynamical process of 2019-nCoV spreading, and based on the collected data of 2019-nCoV, the basic reproduction number is predicted in this article a little higher than the SARS, which suggests that 2019-nCoV is of a moderate-high transmissibility. In order to control the further extension of 2019-nCoV quickly, more effective and timely control measures are required. In the process of prediction, some parameters related to the early stage of SARS transmission were used, and meanwhile it shows that the basic reproduction number is a little sensitive to the generation time. Thus, the microscopic survey about the transmission processes is needed to further improve the quality of predictions.

**Notes Added**. Two days after uploading the first version to arXiv, we are aware of a newly online published article [23] that has directly estimated some epidemiological determinants based on the 425 early confirmed cases in Wuhan. In particular, they suggested the mean serial interval as 7.5 days (95%CI 5.3 to 19), and the mean incubation period as 5.2 days (95%CI 4.1 to 7.0). Taking their data as reference (since some infections may be unreported, their estimated number also suffers possible bias, but we believe it should be more accurate than using the empirical values from SARS), the ratio should be $\rho = 0.693$ instead of $\rho = 0.65$ (we still assume that the individual infectivity in the incubation period is much smaller than that in the infected period), and the serial interval should be $T_g = 7.5$ while the values we considered in this work is larger, as $T_g = 8.4$ and $T_g = 10.0$. If we still use the People's Daily Reports, say $Y(48)=2820$, then the predicted basic reproduction number is $R_0$=2.2, and if we use the Northeastern University Reports, say $Y(43)=4050$ and $Y(47)=12700$, then the predicted basic reproduction numbers are $R_0$=2.9 and $R_0$=3.0. The worst case we considered in the parameter sensitivity analysis, say $Y(48)=28016$, leads to $R_0$=3.1.

Therefore, using the updated data for 2019-nCoV, we predict the basic reproduction number is in between 2.2 and 3.0. Notice that, the updated prediction is smaller than our previous prediction mainly because of a shorter serial interval. However, this smaller number of $R_0$ does not indicate a more controllable situation because shorter serial interval at the same time means faster propagation.

**References**

1 Adnerson RM, May RM, Infectious Diseases of Humans: Dynamics and Control. Oxford University Press, Oxford, 1991.
2 Available from: https://3g.dxy.cn/newh5/view/pneumonia?scene=2&clicktime=1579583352&enterid=1579583352&from=timeline&isappinstalled=0
3 Available from: https://m.weibo.cn/u/2803301701.
4 Chinazzi M, Davis JT, Gioannini C, *et al*. Series Reports Entitled "Preliminary assessment of the International Spreading Risk Associated with the 2019 novel Coronavirus (2019-nCoV) outbreak in Wuhan City" (unpublished).
5 Donnelly CA, Ghani AC, Leung GM, *et al*. Epidemiological determinants of spread of causal agent of severe acute respiratory syndrome in Hong Kong. Lancet 361, 2003: 1761-1766.
6 Chan JFW, Yuan S, Kok KH, *et al*. A familial cluster of pneumonia associated with the 2019 novel coronavirus indicating person-to-person transmission: a study of a family cluster. Lancet (in press).
7 Pastor-Satorras R, Castellano C, Mieghem PV, *et al*. Epidemic processes in complex networks. Review of Modern Physis 2015, 87(3): 925-979.
8 Wallinga J, Lipsitch M, How generation intervals shape the relationship between growth rates and reproductive numbers. Proceedings of the Royal Society B: Biological Sciences, 2007, 274(1609): 599-604.
9 Imai N, Dorigatti I, Cori A, *et al*. Series Reports Entitled "Estimating the potential total number of novel Coronavirus in Wuhan City, China" (unpublished).
10 Read JM, Bridgen JRE, Cummings DAT, *et al*. Novel coronavirus 2019-nCoV: early estimation of epidemiological parameters and epidemic predictions. Preprint in MedRXiv 2020 (unpublished).
11 Leo YS, Chen M, Heng BH, *et al*. Severe acute respiratory syndrome-Singapore 2003. Morb. Mortal. Weekly Report, 2003, 52(18): 405-432.
12 Lipsitch M，Cohen T, Cooper B, *et al*. Transmission Dynamics and Control of Severe Acute Respiratory Syndrome. Science, 2003, 300(5627): 1966-1970.
13 Imai N, Cori A, Dorigatti I, *et al*. Transmissibility of 2019-nCoV (unpublished).
14 Riley S. Transmission Dynamics of the Etiological Agent of SARS in Hong Kong: Impact of Public Health Interventions. Science, 2003, 300(5627): 1961-1966.
15 Wallinga J, Teunis P. Different epidemic curves for severe acute respiratory syndrome reveal similar impacts of control measures. American Journal of Epidemiology, 2004, 160(6): 509-517.
16 Lessler J, Chaisson LH, Kucirka LM, *et al*. Assessing the global threat from Zika virus. Science, 2016, 353: aaf8160.
17 Majumder MS, Rivers C, Lofgren E, *et al*. Estimation of MERS-coronavirus reproductive number and case fatality rate for the spring 2014 Saudi Arabia outbreak: insights from publicly available data. PLoS Currents Outbreaks, 2014, 6.
18 Eichner M, Dietz K, Transmission potential of smallpox: estimates based on detailed data from an outbreak, American Journal of Epidemiology, 2003, 158(2): 110-117.

19 Balcan D, Gonçalves B, Hu H, *et al.* Modeling the spatial spread of infectious diseases: The Global Epidemic and Mobility computational model. Journal of Computational Science, 2010, 1(3): 132-145

20 Brockmann D, Helbing D. The Hidden Geometry of Complex, Network-Driven Contagion Phenomena. Science, 2013, 342(6164): 1337-1342.

21 Bengtsson L, Gaudart J, Lu X, *et al.* Using Mobile Phone Data to Predict the Spatial Spread of Cholera. Scientific Report, 2015, 5(8923).

22 Liu QH, Ajelli M, Aleta A, *et al.* Measurability of the epidemic reproduction number in data-driven contact networks. PNAS, 2018, 115(50): 12680-12685.

23 Li Q, Guan X, Wu P, *et al.* Early Transmission Dynamics in Wuhan, China, of Novel Coronavirus–Infected Pneumonia. N. Eng. J. Med. (in press).